\documentclass[acmsmall]{nimeart}
\usepackage{fontspec}
\usepackage{xeCJK}
\setcopyright{cc}
\copyrightyear{2026}
\acmYear{2026}
\acmDOI{}
\acmConference[NIME '26]{International Conference on New Interfaces for Musical Expression}{June 23--26,
  2026}{London, UK}
\acmISBN{}
\begin{document}

\title{回溯：Co-constructing a Dual Feedback Apparatus}

\author{Yichen Wang}
\email{yichen.wang@anu.edu.au}
\affiliation{%
  \institution{The Australian National University}
  \city{Canberra}
  \state{ACT}
  \country{Australia}
}

\author{Charles Patrick Martin}
\email{charles.martin@anu.edu.au}
\orcid{0000-0001-5683-7529}
\affiliation{%
  \institution{The Australian National University}
  \city{Canberra}
  \state{ACT}
  \country{Australia}
}


\maketitle

\begin{figure}[h]
  \centering
  \includegraphics[width=0.85\linewidth]{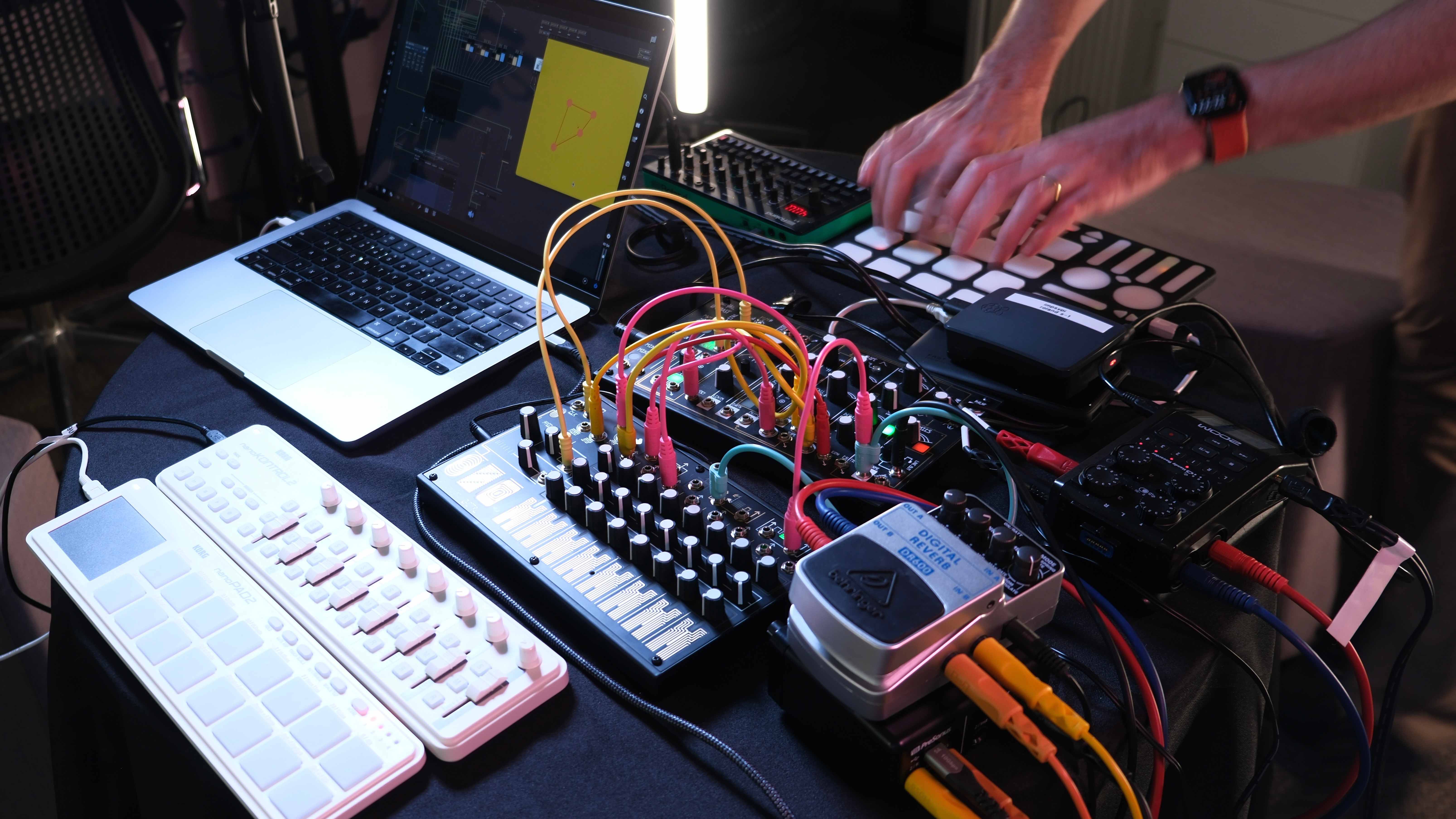}
  \caption{Preparing for a recording with our two intelligent musical instruments which both explore feedback loops with generative AI models. The instruments have generative models that explore feedback in two spaces, audio and control, articulating two approaches to shared musical agency.}
  \label{fig:recording}
\end{figure}

\section{Program Notes}

This performance presents a duet between two intelligent musical instruments 溯 (sù, to trace back; to go upstream) and Agentier (playing on \emph{agentic clavier}), and their human performers, connected through feedback loops.
Rather than treating AI as a tool that responds predictably to input, both systems operate recursively, where past actions continuously influence future behaviour.

The 溯 operates in the audio space through latent representation.
Its performer uses Make Noise 0-series synthesisers and MIDI controllers to work with a neural feedback synthesis system based on a RAVE model, with a latent feedback loop embedded within the model's internal structure. This allows the instrument to remember and reuse its own internal states, influencing ongoing sound generation through its recent sonic history.

The Agentier functions in the control space. Its performer interacts with the system using a Roland S-1 synthesiser and Keith McMillen QuNeo touchpad, where control gestures are routed into a recurrent neural network that feeds back into the synthesis process. 
Through this feedback loop, the system actively shapes the evolution of control signals over time.

Contrasting feedback in the audio and control domains, the performance explores shared agency, resistance, and negotiation between humans and intelligent musical systems. 
Musical phenomena are co-produced through the entangled states of interaction, rather than through pre-existing system configuration or fixed mappings.

\begin{figure}
    \centering
    \includegraphics[width=0.85\linewidth]{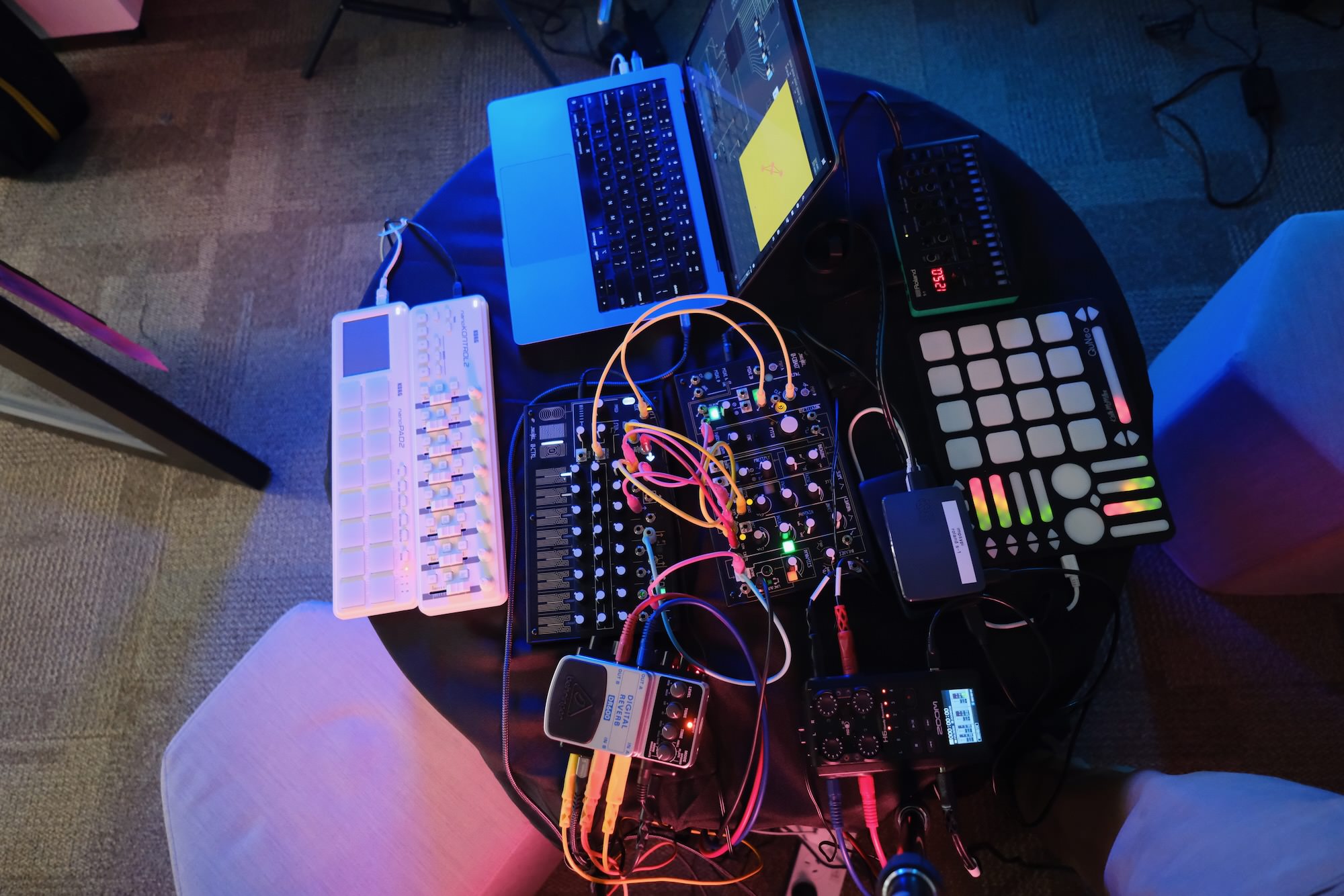}
    \caption{The setup for our dual AI-feedback synthesis performance. Left: Make Noise 0-series synthesisers, laptop and Korg Nano series MIDI controllers, and an external reverb pedal is used as an optional effect during performance. Right: Raspberry Pi, Roland S-1 synthesiser and QuNeo controller.}
    \label{fig:setup-image}
\end{figure}
        
\section{Project Description}

Recent research on musical AI, and particularly the integration of AI algorithms into digital musical instrument (DMI) design (e.g., intelligent musical instruments) has drawn increased attention from musicians toward experimenting with these systems in creative practice~\cite{stefannsdottir2025altered, sophtar:visi2024, livinglooper:shepardson2025, stacco:privato2024, martin_empi}.
These developments open new approaches to understanding the performer-instrument musicianship as a technological and cultural phenomenon~\cite{cultural:jourdan2023}, engaging perspectives from post-humanism~\cite{aihauntography:privato2024} and questioning fundamental principles of new music~\cite{aiterity:tahiroglu2021}.
One approach to this investigation is through material agency, which explores how an instrument actively mediates musical ideas rather than merely transmitting the performer's intention~\cite{mudd2019mateiral}.
In AI-enabled musical instruments, such agency may manifest through data curation and model training~\cite{stefannsdottir2025altered}, feedback interaction~\cite{playingfeedback:mudd2023, kiefer2025feedbackmusicianship}, 
% [graham duuning], expanding composition possibilities through playing on-the-fly, 
and the attribution of autonomous behaviours~\cite{aiterity:tahiroglu2021}.
Through externalising agency, AI-enabled systems form configurations of instrumentality in which the musicians and their instruments enter a mediation where ``the system---along with its generative material traces---contributed to producing a new reality.''~\cite[p. 13]{stefannsdottir2025altered}.

In this performance, we explore feedback musicianship as a means of engaging material agency through two intelligent musical instruments, 溯 and Agentier, each exhibiting distinct forms of feedback.
The duet contrasts two feedback loops operating at different levels of musical organisation, revealing how feedback shapes the ongoing formation of agency in live performance.

\paragraph{溯 (sù, to trace back; to go upstream)}
溯 is a novel latent feedback instrument built on a pretrained RAVE~\cite{rave:caillon2021} vocal model \texttt{Isis}~\cite{ravemodels:acids} implemented in Max/MSP.
The instrument integrates three interacting mechanisms: audio feedback, latent feedback, and direct latent manipulation.
The audio feedback mechanism is introduced as a controllable gain that feeds the model's audio signal back into its input.
A second feedback loop is introduced in RAVE's latent inference process.
% 
% A feedback loop is implemented within the encoder's latent inference process, prior to the decoder for audio reconstruction.
For each latent dimension, the encoder's inferred latent parameters at the current audio window are combined with the corresponding parameters inferred from the previous window.
This latent feedback biases the current latent distribution toward the previous state, introducing temporal dependency
% continuity in the latent space
% By allowing prior latent information to influence the current inference, the system discourages
and reducing abrupt changes between successive windows.
% latent vectors and promotes smoother transitions over time.
% 
In parallel, latent dimensions can be directly manipulated via MIDI control prior to decoding, allowing performers to shape the timbral characteristics of the generated sound. 
These manipulated dimensions remain subject to latent feedback, enabling timbral intervention while preserving temporal continuity.
External hardware synthesisers can also be incorporated as audio input, allowing richer sonic materials to circulate within the feedback structure.

The instrument responds to limitations in neural synthesis systems that rely on learnt latent space representations (e.g., timbre features), which often offer limited control over high-dimensional and abstract sonic features for creative expression~\cite{latentdace:nabi2024}, despite their growing use in performance practice~\cite[e.g.,][]{stacco:privato2024,latentroom:shaheed2024}.
Rather than relying on conditioning or assumptions about learned representations~\cite{latentdace:nabi2024}, 溯 borrows the recursivity of the feedback technique to stabilise navigation in the latent space, supporting more controllable timbral behaviour in performance.

\paragraph{Agentier (agentic clavier)}
Agentier uses a novel musical AI platform to mediate and generate gestural information (MIDI messages) between two performance interfaces. 
An embodied musical gesture AI model using an auto-regressive mixture density recurrent neural network (MDRNN)~\cite{Martin2019} was trained on data from eight continuous controllers, recorded by the second author.
The model can generate and continue streams of musical control data including the time at which updates should be performed.
This model works in a call-and-response loop with the performer with a fast switchover time of 0.1 seconds. In practice, this means that the model can fill in between the performer's gestures and can be guided and mediated by continual interactions.
Two interfaces, a synthesiser and touch-pad controller are connected to the AI model, with input received from and output sent to both simultaneously. The eight control outputs are mapped to notes and seven timbral parameters on the synth and eight LED touch sliders on the touch-pad. Inputs from the touch-pad's sliders and pads, and the synth's knobs and keys map to the eight inputs of the model.
The two interfaces form an assemblage~\cite{musical_instruments_as_assemblage_theberge} with attributes of each interface contributing to the new intelligent instrument.
Situated within recent discussions of Baradian apparatus in DMI and HCI research~\cite{morrison2024entangle, playingfeedback:mudd2023, ambiguity:reed2024}, 
this performance treats performers, intelligent musical instruments, and feedback loops as a coupled configuration. 
Musical phenomena emerge through ongoing intra-action between material, computational, and performative elements, with feedback acting as a generative condition shaping musical behaviour in real time.

\section{Technical Notes}

\begin{figure}
    \centering
    \includegraphics[width=0.75\linewidth]{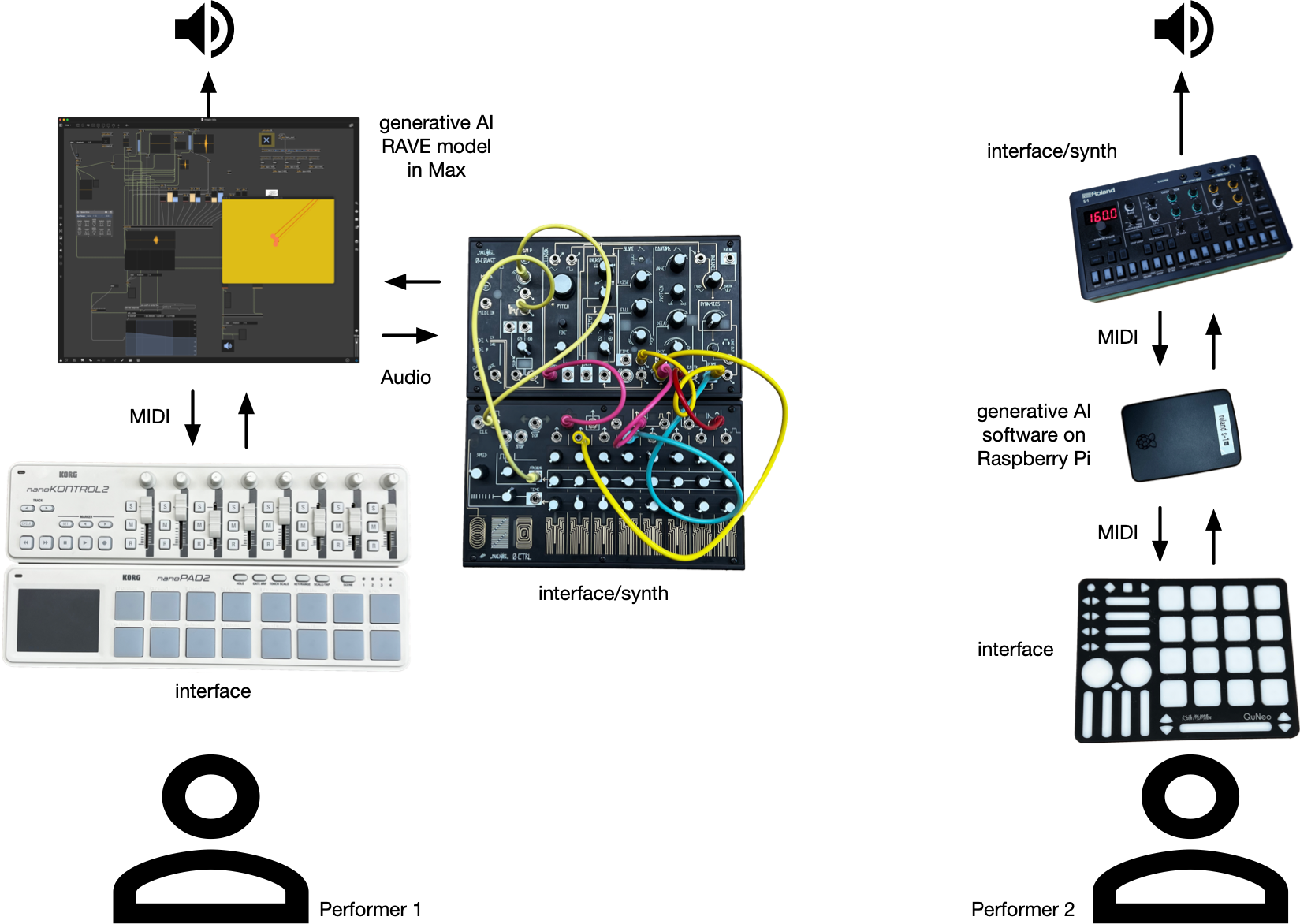}
    \caption{A system diagram for the two intelligent musical instruments in this performance. Performer 1's instrument has a generative audio feedback loop. Performer 2's instrument has a generative MIDI data feedback loop. Both performers share agency with their AI systems, negotiating, resisting and guiding their instruments together to create a musical performance.}
    \label{fig:system-diagram}
\end{figure}

The performance involves two tabletop instrument setups set side-by-side with independent audio output (see figures \ref{fig:setup-image} and \ref{fig:system-diagram}). The following equipment is supplied by the performers:
\begin{itemize}
    \item performer 1: Korg Nano series controllers, Make Noise 0-series synthesisers, reverb pedal, laptop (HDMI output), audio interface (stereo balanced 1/4 inch outputs)
    \item performer 2: QuNeo MIDI controller, Roland S-1 synthesiser, Raspberry Pi (stereo 3.5mm output)
    \item overhead camera with HDMI output (GoPro)
\end{itemize}
The following equipment is required at the venue:
\begin{itemize}
    \item trestle table to fit both performers
    \item two chairs
    \item two stereo DIs (4 channels of DI in total)
    \item HDMI inputs for laptop and overhead camera
    \item 6 power outlets
\end{itemize}

\section{Media Links}

\begin{itemize}
	\item Video: \url{https://doi.org/10.5281/zenodo.19673150} (also included as supplementary material).
\end{itemize}
This video documents a 12-minute improvised performance, recorded following a process of artistic development and two live concerts. We envision a similar performance at NIME.

\section{Ethical Standards}

The instrument 溯 was developed by the first author during a research internship program at~\anon{Dolby ATG Group, Sydney, Australia} in 2025.
This work does not involve human participants, animal subjects, or the collection of personal data.
Aspects of the instrument 溯 are subject to a patent application with~\anon{Dolby ATG}.

\begin{acks}
The first author would like to thank David Cooper, Benjamin Southwell, David Gunawan and Jeroen Breebaart for their supervision and guidance in the development of the instrument 溯 in her research and musical practices.
\end{acks}

\bibliographystyle{ACM-Reference-Format}
\bibliography{references}
\end{document}